\documentclass[a4paper,11pt]{article}
\usepackage{pos}

\usepackage{graphicx}
\usepackage{graphics}
\usepackage{multirow}
\usepackage{multicol}
\usepackage{caption}
\usepackage{subcaption}
\usepackage{comment}

\usepackage{amsfonts}
\usepackage{amsmath}
\usepackage{mathtools}

\DeclareMathAlphabet{\mathcal}{OMS}{cmsy}{m}{n}
\newcommand{\Nf}{{N_f}}
\newcommand{\psibar}{{\bar{\psi}}}
\newcommand{\corr}[1]{\langle #1\rangle}
\newcommand{\mathsp}{\;\;\;\;}

\newcommand{\bs}[1]{\boldsymbol{#1}}

\newcommand{\sss}{\scriptscriptstyle}

\newcommand{\Zgs}[1][6]{Z_G^{\{#1\}}}

\newcommand{\Zfs}[1][6]{Z_F^{\{#1\}}}

\newcommand{\Tgs}[1][\mu\nu]{T^{G,\{6\}}_{#1}} 
\newcommand{\Tgt}[1][\mu\nu]{T^{G,\{3\}}_{#1}}
\newcommand{\Tfs}[1][\mu\nu]{T^{F,\{6\}}_{#1}}
\newcommand{\Tft}[1][\mu\nu]{T^{F,\{3\}}_{#1}}

\newcommand{\NuAB}{{{\cal V}_{0,k}^{AB}}}

\newcommand{\vxi}{\boldsymbol{\xi}}

\newcommand{\vxithzero}[1]{ { {\sss\vxi,}{\sss\theta_{\scalebox{0.50}{0}}^{#1}} } }

\newcommand{\LambdaQCD}{\Lambda_{\text{QCD}}}

\title{Progresses on high-temperature QCD: Equation of State and energy-momentum tensor}
\ShortTitle{Progresses on high-temperature QCD: Equation of State and energy-momentum tensor}

\author*[a,b]{Matteo Bresciani}
\author[c]{Mattia Dalla Brida}
\author[a,b]{Leonardo Giusti}
\author[b]{Michele Pepe}

\affiliation[a]{Dipartimento di Fisica, Università di Milano-Bicocca,\\
Piazza della Scienza 3, I-20126 Milano, Italy}

\affiliation[b]{INFN, Sezione di Milano-Bicocca,\\
Piazza della Scienza 3, I-20126 Milano, Italy}

\affiliation[c]{Theoretical Physics Department, CERN\\
  1211 Geneva 23, Switzerland}

\emailAdd{Matteo.Bresciani@mib.infn.it}
\emailAdd{mattia.dalla.brida@cern.ch}
\emailAdd{Leonardo.Giusti@mib.infn.it}
\emailAdd{Michele.Pepe@mib.infn.it}

\abstract{
We present first non-perturbative results for the renormalization constants of the QCD energy-momentum tensor, based on the framework of thermal QCD with shifted and twisted (for quarks only) boundary conditions in the compact direction. We also show preliminary results for the entropy density obtained with the very same numerical strategy. This opens the way to the determination of the QCD Equation of State up to very high temperatures.

\begin{flushright}
CERN-TH-2023-219
\end{flushright}
}

\FullConference{%
The 40th International Symposium on Lattice Field Theory,\\
July 31st-August 4th  2023,\\
Fermilab, Batavia, Illinois, USA
}

\begin{document}
\maketitle

\numberwithin{equation}{section}

\section{Introduction}
\label{sec:section1}
Quantum Chromodynamics (QCD) is the fundamental field theory describing the interaction among quarks and gluons. 
Its thermal state from the deconfinement temperature $150$-$200$ MeV up to the electroweak scale $\sim100$ GeV, the so-called Quark-Gluon Plasma (QGP),
is of great interest for instance in cosmology and high energy physics. Indeed the Big Bang model describes the Universe in some of its early stages as a plasma of quarks and gluons.
Moreover, the QGP is currently under investigation in heavy-ion collision experiments. 
On the theoretical side, many physical properties of the QGP are encoded by the correlation functions of the energy-momentum tensor (EMT) of
QCD, see Ref.~\cite{Meyer:2011gj} for a recent review. They  can be computed from first principles on
the lattice once the discretized EMT has been properly renormalized. We propose a numerical strategy to determine the renormalization constants of the EMT non-perturbatively,
based on considering QCD with shifted and twisted boundary conditions~\cite{Giusti:2012yj,DallaBrida:2020gux}.
The very same framework allows for the determination of the Equation of State (EoS) of QCD up to very high temperatures.
At present the EoS is known non-perturbatively for temperatures below $1$ GeV~\cite{HotQCD:2014kol,Borsanyi:2013bia} because of some
technical limitations of the state of the art techniques, which are overcome by our alternative strategy. An analogous approach has been
exploited for the SU$(3)$ Yang-Mills theory in Ref.~\cite{Giusti:2016iqr}, where perturbation theory was shown to be unreliable up to very high temperatures.

\section{The energy-momentum tensor in the continuum}
\label{sec:section2}
We consider the symmetric and gauge-invariant definition of the QCD EMT~\cite{Landau:1975pou, Caracciolo:1989pt}. In the continuum the latter splits
into a singlet and a 9-dimensional representation of the SO$(4)$ group. Under renormalization, the singlet component mixes with the identity operator, while there are
no other fields of mass dimension $\leq4$ which may mix with the non-singlet representation. Moreover, the bare and renormalized non-singlet EMT coincide because the latter is a conserved current associated to the invariance of the theory under the group of space-time translations. This property becomes manifest if we formulate QCD at finite temperature $T$, in a moving reference frame at Euclidean speed $\vxi$ and with an imaginary chemical potential $\mu_I$~\cite{Giusti:2012yj,DallaBrida:2020gux}. This setup is realized with the following boundary conditions in the (compact) temporal direction for the gauge and quark fields~\cite{DallaBrida:2020gux}:
\begin{equation*}
    A_\mu(x_0+L_0,\bs{x}) = A_\mu(x_0,\bs{x}-L_0\bs{\xi})\;,
\end{equation*}
\begin{equation}
    \psi(x_0+L_0,\bs{x}) = -e^{i \theta_0}\, \psi(x_0,\bs{x}-L_0\bs{\xi})\;,
    \label{eq:shtwbcs}
\end{equation}
\begin{equation*}
    \psibar(x_0+L_0,\bs{x}) = -e^{-i \theta_0}\, \psibar(x_0,\bs{x}-L_0\bs{\xi})\;.
\end{equation*}
The Euclidean boost is encoded by the $L_0\vxi$ shift ($L_0$ is the temporal extension) in the spatial directions at the temporal boundary. The usual relation between the temperature and $L_0$ is modified by the Euclidean Lorentz factor in $T = L_0^{-1}(1+\vxi^2)^{-1/2}$, and the twist phase $\theta_0=-L_0\mu_I$ for the fermions is related to the imaginary chemical potential. 
The space-time components of the EMT satisfy the following identity~\cite{DallaBrida:2020gux},
\begin{equation}
    \corr{T_{0k}}_\vxithzero{} = -\dfrac{\partial f_\vxithzero{}}{\partial\xi_k}\;, \quad f_\vxithzero{} = -\frac{1}{L_0L^3}\ln {\cal Z}(L_0, L, \vxi, \theta_0)\;,
    \label{eq:Tfen}
\end{equation}
where $f_\vxithzero{}$ is the free-energy density of thermal QCD and $\mathcal{Z}$ its partition function. The free-energy density is a spectral quantity with an additive power-divergent term. Its derivative with respect to the shift $\vxi$ is thus finite, once the bare parameters of the theory have been renormalized. Therefore $T_{0k}$ and all the other non-singlet components of the EMT are finite too. The traceless diagonal components of the EMT are related to the space-time components by means of Lorentz transformations~\cite{Giusti:2012yj,DallaBrida:2020gux}:
\begin{equation}
    \corr{T_{0k}}_\vxithzero{} = \xi_k \left(\corr{T_{00}}_\vxithzero{}-\corr{T_{jj}}_\vxithzero{}\right) \quad (j\neq k, \xi_j=0)\;.
    \label{eq:T0kTjj}
\end{equation}
Finally, the dependence of $\corr{T_{0k}}_\vxithzero{}$ on the twist phase $\theta_0$ can be written as~\cite{DallaBrida:2020gux}
\begin{equation}
    \corr{T_{0k}}_\vxithzero{A} - \corr{T_{0k}}_\vxithzero{B} =
      \dfrac{i}{L_0}\int_{\theta_0^A}^{\theta_0^B}d\theta_0\dfrac{\partial}{\partial\xi_k}\corr{V_0}_\vxithzero{} 
      \equiv - \NuAB \;,
      \label{eq:NuAB}
\end{equation}
where $V_\mu$ is the flavour-singlet vector current. We will take advantage of these relations in our lattice calculations.
\section{Renormalization of the energy-momentum tensor on the lattice}
\label{sec:section3}
The lattice regularization breaks the SO$(4)$ symmetry down to its discrete hypercubic subgroup SW$_4$. The non-singlet EMT components split into
the sextet and triplet representations of this discrete group~\cite{Caracciolo:1989pt}. These two representations renormalize separately, and no other lattice
field is involved in their renormalization. Since only discrete space-time translations are allowed, the bare non-singlet EMT is not a conserved current.
As a consequence, the renormalization constants of the operators in each representation are finite functions of the bare coupling $g_0$ and independent from the external states, up to discretization
effects. We focus on the non-perturbative definition of the sextet and triplet representations of the lattice EMT, which follow the renormalization pattern
\begin{equation}
    T_{\mu\nu}^{R,\{i\}} = Z_G^{\{i\}} T_{\mu\nu}^{G,\{i\}} + Z_F^{\{i\}} T_{\mu\nu}^{F,\{i\}}, \mathsp i=3,6 \;.
    \label{eq:defZT}
\end{equation}
$T_{\mu\nu}^{G,\{i\}}$ and $T_{\mu\nu}^{F,\{i\}}$ are the dimension-4 gauge-invariant lattice operators in the $i-$dimensional representation of the hypercubic group~\cite{Caracciolo:1989pt,DallaBrida:2020gux}.

The lattice shifted and twisted boundary conditions are analogous to the continuum ones \eqref{eq:shtwbcs}, but with the gauge potential $A_\mu(x)$ replaced by the link field $U_\mu(x)$. In addition, we consider the usual periodic boundary conditions for the three compact spatial directions of the lattice box, each of size $L$.

Using the definition \eqref{eq:defZT} together with equation \eqref{eq:Tfen} at two values of the twist phase, $\theta_0^A$ and $\theta_0^B$, and exploiting equation \eqref{eq:NuAB}, we can write our master formula for the computation of the renormalization constants of the sextet representation:
\begin{gather}
\begin{pmatrix}
    \corr{\Tgs[0k]}_\vxithzero{A} & & \corr{\Tfs[0k]}_\vxithzero{A} \\ \\ \\
    \corr{\Tgs[0k]}_\vxithzero{B} & & \corr{\Tfs[0k]}_\vxithzero{B}
\end{pmatrix}
\begin{pmatrix}
    \Zgs \\ \\ \\
    \Zfs
\end{pmatrix}
=
\begin{pmatrix}
    - \dfrac{\Delta f_\vxithzero{A}}{\Delta\xi_k}\\ \\
    - \dfrac{\Delta f_\vxithzero{A}}{\Delta\xi_k} + \NuAB
\end{pmatrix}
+ {\rm O}(a^2) \;.
\label{eq:master}
\end{gather}
The partial derivative has been replaced by finite differences up to discretization effects. Then, the lattice counterpart of the Ward Identity \eqref{eq:T0kTjj}
\begin{equation}
        \corr{T^{R,\{6\}}_{0k}}_\vxithzero{} = \xi_k\corr{T^{R,\{3\}}_{0j}}_\vxithzero{}
        \mathsp (j\neq k, \xi_j=0) \;
        \label{eq:63}
\end{equation}
gives access to the renormalization constants of the triplet representation. By dimensional arguments, the leading lattice artifacts of the renormalization constants when $a\to 0$ are 
\begin{equation}
    Z \left(g_0^2, aT\right) = Z(g_0^2) + C_1 \cdot \left(aT\right)^2 + C_2\cdot(a\LambdaQCD)aT + C_3\cdot(a\LambdaQCD)^2 + \;... \;,
\end{equation}
where $Z$ is a shorthand for any of the renormalization constants in equation \eqref{eq:defZT}. We include a $T\to 0$ limit at fixed $g_0^2$ in the definition of the renormalization constants, so that the leading discretization effects $C_1$, $C_2$ are suppressed. All the residual lattice artifacts will disappear when a renormalized correlator of the lattice EMT will be extrapolated to the continuum limit.

\section{Renormalization strategy at work}
\label{sec:section4}
This procedure proved to be effective in the non-perturbative renormalization of the energy-momentum tensor in the SU$(3)$ Yang-Mills theory on the lattice~\cite{Giusti:2015daa}, and of the flavour-singlet local vector current~\cite{Bresciani:2022lqc} in lattice QCD. In the following we give some details on the lattice determination of the quantities appearing in the master equation \eqref{eq:master}. Although the strategy is general, we choose the Wilson formulation of lattice QCD with $\Nf=3$ flavours of massless O($a$)-improved Wilson fermions.

\subsection{Shift derivative of the free-energy density}
\label{ssec:integral_mass}
In order to estimate the shift derivative of the free-energy from a lattice calculation we add and subtract inside the discrete differential operator, at fixed $g_0^2$ and $L_0/a$, the free-energy density $f^\infty_\vxithzero{}$ of QCD with quarks at infinite mass:
\begin{equation}
    \dfrac{\Delta f_\vxithzero{}}{\Delta\xi_k}
     = \dfrac{\Delta}{\Delta\xi_k}\left(f_\vxithzero{}^\infty- \int_0^{\infty}dm_q\;\dfrac{\partial f_\vxithzero{}}{\partial m_q} \right)
     = \dfrac{\Delta}{\Delta\xi_k}\left(f_\vxithzero{}^\infty - \int_0^{\infty}dm_q\;\corr{\psibar\psi}_\vxithzero{} \right)\;.
    \label{eq:Df}
\end{equation}
We also replace $f_\vxithzero{} - f^\infty_\vxithzero{}$ with an integral in the bare subtracted quark mass $m_q=m_0-m_c$ ($m_c$ is the critical mass) of the derivative in $m_q$ of $f_\vxithzero{}$. This derivative is the chiral condensate, as we write in the last step. The integral in the mass is a well defined quantity because the additive divergences of the chiral condensate are canceled by the overall differentiation in the shift, while its multiplicative renormalization drops together with the renormalization of the quark mass in the integration measure.

\begin{figure}
     \centering
     \begin{subfigure}[b]{0.5\textwidth}
         \centering
         \includegraphics[width=\textwidth, trim=0.cm .3cm 0.cm 0.cm, clip]{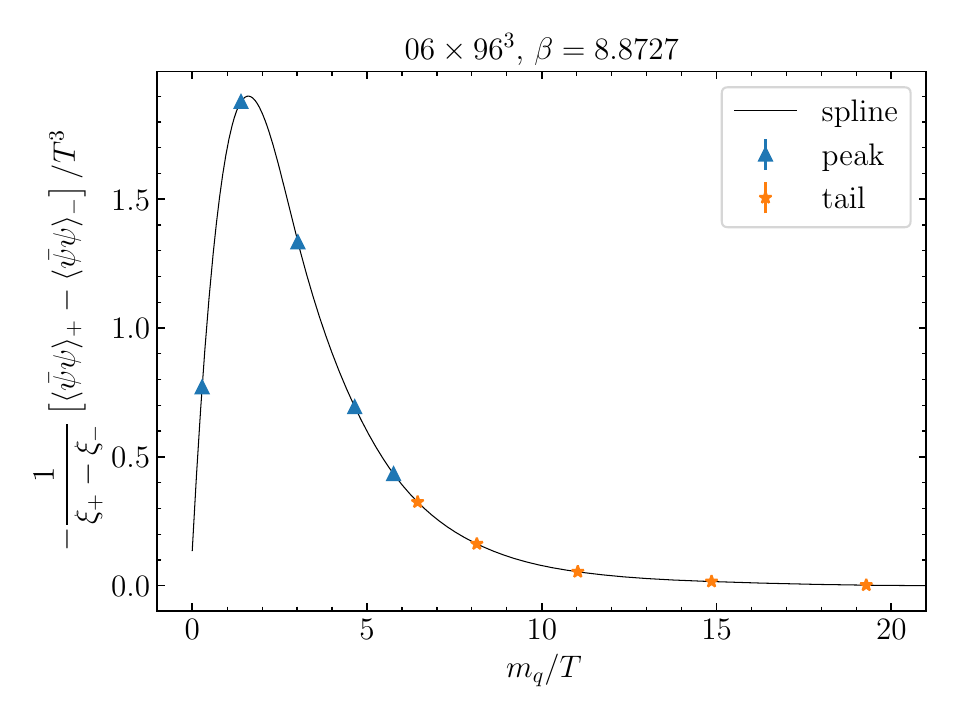}
     \end{subfigure}%
     \begin{subfigure}[b]{0.5\textwidth}
         \centering
         \includegraphics[width=\textwidth]{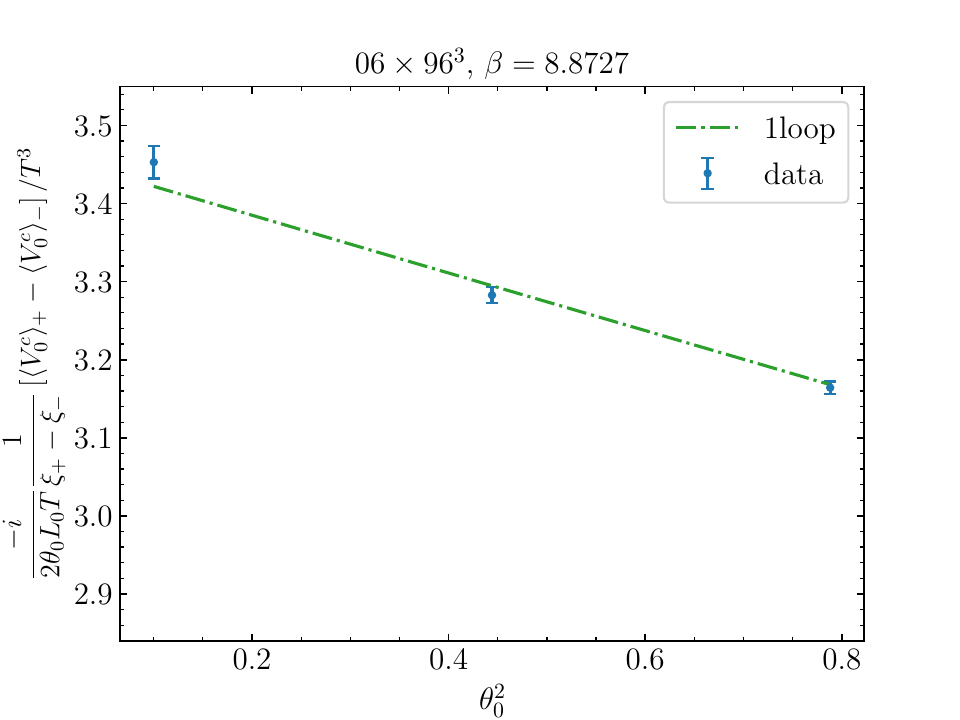}
     \end{subfigure}%
     \caption{Left: integrand function of eq. \eqref{eq:Df} sampled at the mass values prescribed by the Gauss quadratures for peak and tail.  Right: integrand function of eq. \eqref{eq:Nu_lat} sampled at the values of the twist phase $\theta_0$ prescribed by the Gauss quadrature.}
     \label{fig:integrals}
\end{figure}

Quarks at infinite mass decouple from QCD, therefore the shift derivative of $f^\infty_\vxithzero{}$ can be computed in the $\Nf=0$ theory as described in~\cite{Giusti:2015daa}. We estimate the integral in the quark mass in lattice QCD, and the left panel of Figure \ref{fig:integrals} shows a preliminary plot of the integrand function coming from lattice simulations on a $6\times96^3$ volume and at $\beta=8.8727$. The integral is estimated with two Gauss quadratures, one for the peak (5 points) and one for the tail (10 points). The integration interval is cut when the remainder of the integral is expected to be negligible compared with the target accuracy. Using this strategy we obtain the area below the curve with a relative precision of a few permille.

\subsection{Integral in the twist phase}
We move to the determination of the integral $\NuAB$ in equation \eqref{eq:master}, defined in equation \eqref{eq:NuAB}. On the lattice we have
\begin{equation}
    \NuAB = -\dfrac{i}{L_0} \int_{\theta_0^A}^{\theta_0^B} d\theta_0\; \dfrac{\Delta}{\Delta \xi_k} \corr{V_0^R}_\vxithzero{} 
    = -\dfrac{i}{L_0}\int_{(\theta_0^A)^2}^{(\theta_0^B)^2}d\theta_0^2\;\frac{1}{2\theta_0}\dfrac{\Delta}{\Delta \xi_k} \corr{V_0^R}_\vxithzero{} \;,
    \label{eq:Nu_lat}
\end{equation}
and in place of $V_0^R$ we may use the conserved current $V_0^c$, or the renormalized local current $Z_VV_0^l$. For this study, we choose the former. As a last technical step, we perform a change of variable from $\theta_0$ to $\theta_0^2$, and we estimate the resulting integral with a $n=3$ Gauss quadrature. The change of variable is convenient because the integrand function is odd in $\theta_0$, meaning that a polynomial representation of it, like the Gauss quadrature, would contain odd powers of $\theta_0$ only. With the change to $\theta_0^2$ we effectively halve the degree of the polynomial, so that with a 3-point quadrature we obtain the same accuracy that we would get with a 6-point quadrature of the original integral. The interplay of the twist phase with the $\mathbb{Z}_3$ center symmetry of the gauge group SU$(3)$ makes the free-energy density effectively periodic in $\theta_0$, with period $2\pi/3$~\cite{Roberge:1986mm}. The choice $\theta_0^A=0$ and $\theta_0^B=3\pi/10$ covers the most of the allowed interval and thus maximizes the signal. The right panel of Figure \ref{fig:integrals} shows the integrand function, compared with its 1-loop result at finite $L_0/a$. The lattice setup is the same as for the integral in the mass, and as in that case we obtain a relative error of a few permille on the integral.

\subsection{1-point functions of the bare EMT}
The last missing ingredients in equation \eqref{eq:master} are the 1-point functions of the bare EMT appearing in the matrix on the left-hand side. Table \ref{tab:tmunu} shows some preliminary numerical results at the usual bare parameters. Data at $\theta_0=0$ come from ensembles with $L/a=288$ while data at $\theta_0=3\pi/10$ from ensembles with $L/a=96$. Finite volume effects are expected to be
negligible because exponentially suppressed with the mass of the lightest screening state~\cite{Giusti:2012yj}, which is proportional to the temperature. The 1-point functions are averaged over the volume, and the number of measurements is tuned so as to reach a comparable statistical precision in the two ensembles. The relative error on the $\corr{T^F}$ components is a few permille, while on the $\corr{T^G}$ components is about $1\%$. 

\begin{table}
    \centering
    \begin{tabular}{ccc}
        \hline
        \multicolumn{3}{c}{$\theta_0^A=0$}\\
        \hline
        $\corr{\Tfs[01]}/T^4$ & -6.343(9) & 0.15$\%$ \\ \rule{0pt}{4ex}
        $\corr{\Tgs[01]}/T^4$ & -2.822(24) & 0.85$\%$ \\ \rule{0pt}{4ex}
        $\corr{\Tft[02]}/T^4$ & -6.846(11) & 0.17$\%$ \\ \rule{0pt}{4ex}
        $\corr{\Tgt[02]}/T^4$ & -3.13(4) & 1.20$\%$ \\
        \hline
        \multicolumn{3}{c}{$L/a=288$, $n_{\rm trj}=100$}\\
    \end{tabular}
    \centering
    \quad
    \begin{tabular}{ccc}
        \hline
        \multicolumn{3}{c}{$\theta_0^B=3\pi/10$}\\
        \hline
        $\corr{\Tfs[01]}/T^4$ & -4.053(9) & 0.21$\%$ \\ \rule{0pt}{4ex}
        $\corr{\Tgs[01]}/T^4$ & -2.677(27) & 1.00$\%$ \\ \rule{0pt}{4ex}
        $\corr{\Tft[02]}/T^4$ & -4.375(15) & 0.34$\%$ \\ \rule{0pt}{4ex}
        $\corr{\Tgt[02]}/T^4$ & -2.86(4) & 1.53$\%$ \\
        \hline
        \multicolumn{3}{c}{$L/a=96$, $n_{\rm trj}=2000$}\\
    \end{tabular}
    \caption{Preliminary results for the 1-point functions of the sextet and triplet bare EMT. The bare parameters are $L_0/a=6$, $\beta=8.8727$.}
    \label{tab:tmunu}
\end{table}

\subsection{Renormalization constants}
\begin{figure}
     \centering
     \begin{subfigure}[b]{0.5\textwidth}
         \centering
         \includegraphics[width=\textwidth, trim=2.5cm 0.cm 0.cm 1.cm, clip]{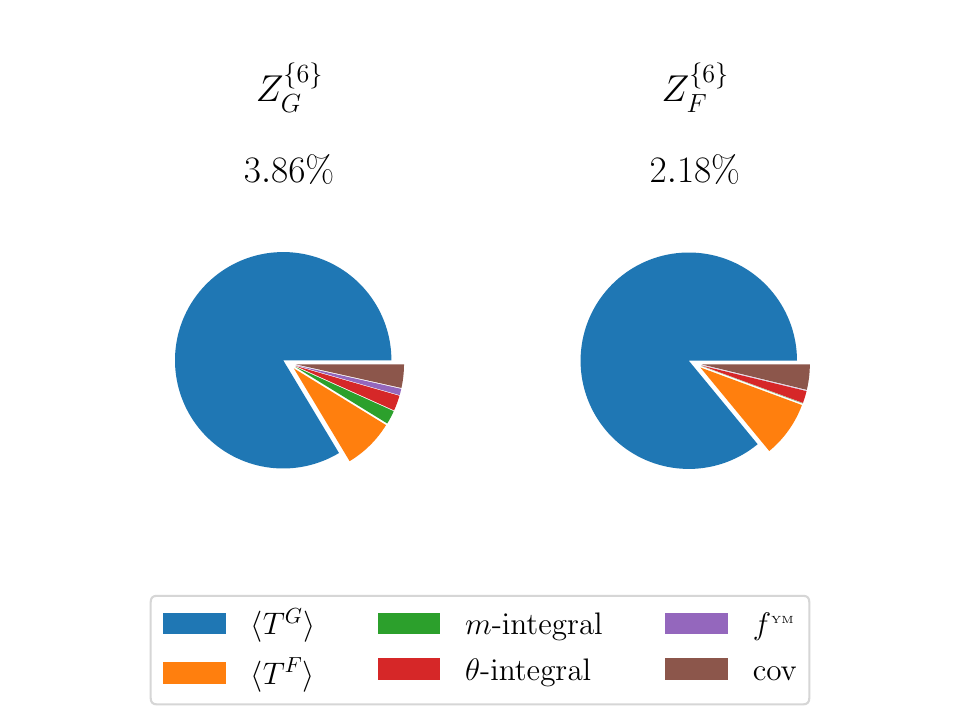}
     \end{subfigure}%
     \begin{subfigure}[b]{0.5\textwidth}
         \centering
         \includegraphics[width=\textwidth, trim=2.5cm 0.cm 0.cm 1.cm, clip]{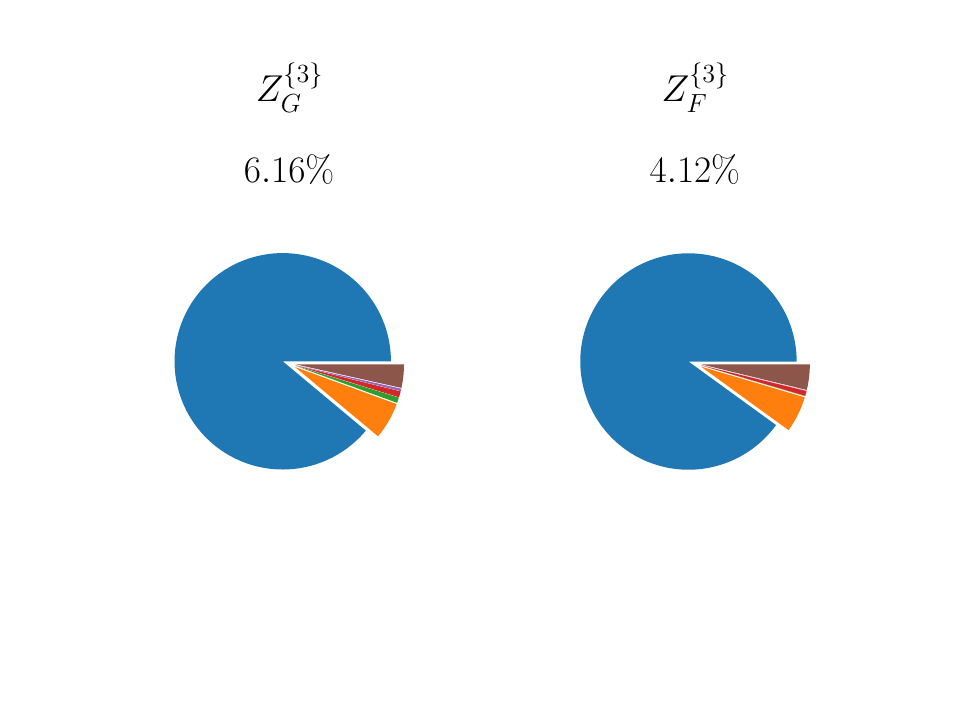}
     \end{subfigure}%
     \caption{Breakdown of the contributions to the variance of the renormalization constants.}
     \label{fig:pie}
\end{figure}

Using the results described so far in equation \eqref{eq:master} we obtain a first estimation of the renormalization constants of the sextet energy-momentum tensor. Then, equation \eqref{eq:63} leads to the renormalization constants of the triplet representation. Figure \ref{fig:pie} shows the relative errors on the four renormalization constants, and the pie charts give the breakdown of the contributions to the variances. The error is largely dominated by the one of the 1-point functions $\corr{T^G}$. The reason can be traced back to the fact that the solution of the master equation \eqref{eq:master} depends on differences like $\corr{T^G}_\vxithzero{B}-\corr{T^G}_\vxithzero{A}$. These 1-point functions are weakly dependent on $\theta_0$, therefore the signal mostly cancels while the errors, which are basically independent from the value of $\theta_0$, sum in quadrature.

\section{Entropy density}
\label{sec:section5}
The numerical determination of the shift derivative of the free-energy density is closely related to the computation of the entropy density of QCD~\cite{Giusti:2012yj,DallaBrida:2020gux}: 
\begin{equation}
    \dfrac{s}{T^3} = \dfrac{1+\vxi^2}{\xi_k}\dfrac{1}{T^4}\dfrac{\Delta f_{\vxi}}{\Delta \xi_k} 
    = \dfrac{1+\vxi^2}{\xi_k}\dfrac{1}{T^4}\left(\dfrac{\Delta f^\infty_{\vxi}}{\Delta \xi_k}-\int_0^\infty dm_q \dfrac{\Delta\corr{\psibar\psi}_{\vxi}}{\Delta\xi_k}\right)\;.
\end{equation}
The free-energy is expressed in terms of the infinite quark mass case and of an integral in the quark mass of the shift derivative of the chiral condensate, similarly to what described in Subsection \ref{ssec:integral_mass}. Using these formulas, and the lattice data at $L_0/a=6$, $\beta=8.8727$ (corresponding to a temperature of about $140$ GeV~\cite{DallaBrida:2021ddx}) we estimate the entropy density with a relative error of half a percent. After the tree-level improvement of the lattice data (which is expected to remove the leading discretization effects at small $g_0^2$) we observe a deviation of a few percent with respect to the Stefan-Boltzmann limit.

\section{Conclusions}
\label{sec:section6}
In these proceedings we described our ongoing work on the non-perturbative renormalization of the QCD energy-momentum tensor, and on the first principle determination of the Equation of State of the Quark-Gluon Plasma at high temperature. We carry out our computations in the framework of QCD in a moving reference frame and with an imaginary chemical potential, where some convenient Ward Identities can be employed to constrain the renormalization constants of the energy-momentum tensor, and to compute the entropy density. We showed some preliminary non-perturbative results from a lattice setup of temporal extension $L_0/a=6$ and $\beta=8.8727$, where we obtained a final accuracy of $\sim3\%$ on the renormalization constants $Z_{F}^{\{3,6\}}$, and of $\sim5\%$ on $Z_{G}^{\{3,6\}}$. On the same ensembles, the entropy density is determined with a relative error of $\sim0.5\%$. These results are encouraging for the extension of this computation to the temperature range $1$ GeV - $100$ GeV considered in Ref.~\cite{DallaBrida:2021ddx}.

\section{Acknowledgements}
We acknowledge PRACE for awarding us access to the HPC system MareNostrum4 at the Barcelona Supercomputing Center (Proposals n. 2018194651 and 2021240051) where most of the numerical results presented in this paper have been obtained. We also thank CINECA for providing us with computer time on Marconi (CINECA-INFN, CINECA-Bicocca agreements). The R\&D has been carried out on the PC clusters Wilson and Knuth at Milano-Bicocca. We thank all these institutions for the technical support. This work is (partially) supported by ICSC – Centro Nazionale di Ricerca in High Performance Computing, Big Data and Quantum Computing, funded by European Union – NextGenerationEU. MB would like to thank CERN for the kind hospitality and support extended to him during the write-up of this contribution.

\end{document}